\documentclass[]{spie}  

\usepackage{graphicx} 
\usepackage{amsmath,amsfonts,amssymb}
\usepackage{multirow}
\usepackage{caption}
\usepackage{subcaption}
\usepackage[normalem]{ulem}
\useunder{\uline}{\ul}{}
\usepackage[colorlinks=true, allcolors=blue]{hyperref}

\title{Improving polarimetric accuracy of RoboPol to $<$ 0.05~\% using a half-wave plate calibrator system}

\author[a,b]{Siddharth Maharana}
\author[b,c]{Dmitry Blinov}
\author[b,d,e]{A. N. Ramaprakash}
\author[b,c]{Vasiliki Pavlidou}
\author[b,c]{Konstantinos Tassis}
\affil[a]{South African Astronomical Observatory, PO Box 9, Observatory, 7935, Cape Town, South Africa}
\affil[b]{Institute of Astrophysics, Foundation for Research and Technology - Hellas, Vasilika Vouton, GR-70013 Heraklion, Greece}
\affil[c]{Department of Physics, University of Crete, Voutes University Campus, GR-70013 Heraklion, Greece}
\affil[d]{Inter-University Centre for Astronomy and Astrophysics, Post Bag 4, Ganeshkhind, Pune - 411 007, India }
\affil[e]{Cahill Center for Astronomy and Astrophysics, California Institute of Technology, Pasadena, CA, 91125, USA}

\authorinfo{Further author information: (Send correspondence to S.M.)\\S.M: E-mail: siddharth@saao.ac.za, sidh345@gmail.com}

\begin{document} 
\maketitle

\begin{abstract}
\par RoboPol is a four-channel, one-shot linear optical polarimeter that has been successfully operating since 2013 on the 1.3 m telescope at Skinakas Observatory in Crete, Greece. Using its unique optical system, it measures the linear Stokes parameters $q$ and $u$ in a single exposure with high polarimetric accuracy of 0.1\%-0.15\% and 1 degree in polarization angle in the R broadband filter. Its performance marginally degrades in other broadband filters.

\par The source of the current instrumental performance limit has been identified as unaccounted and variable instrumental polarization, most likely originating from factors such as temperature and gravity-induced instrument flexure. To improve the performance of RoboPol in all broadband filters, including R, we have developed a rotating half-wave plate calibrator system. This calibrator system is placed at the beginning of the instrument and enables modulation of polarimetric measurements by beam swapping between all four channels of RoboPol.

\par Using the new calibrator system, we observed multiple polarimetric standard stars over two annual observing seasons with RoboPol. This has enabled us to achieve a polarimetric accuracy of better than 0.05~\% in both $q$ and $u$, and 0.5 degrees in polarization angle across all filters, enhancing the instrument's performance by a factor of two to three.
\end{abstract}


\section{INTRODUCTION}
\label{sec:intro}  
The Robotic Polarimeter (RoboPol) is a four-channel, one-shot optical linear polarimeter mounted on the 1.3~m telescope at Skinakas Observatory in Crete, Greece\cite{robopol, robopol_pipeline}. It was commissioned in 2013 and has been working successfully since. It is operated by the Institute of Astrophysics, Foundation for Research and Technology-Hellas (FORTH) in Crete, Greece. Since its commissioning, it has been a heavily used instrument. Data obtained by RoboPol have enabled research output in the form of publications in various fields of astronomy, including blazars\cite{robopol_data_release}, gamma-ray bursts\cite{robopol_grb}, and interstellar medium physics\cite{vincent_tomoraphy, Skalidis, gina_extereme_pol}. A comprehensive list of publications resulting from RoboPol data can be found on the RoboPol program's webpage\footnote{\url{https://robopol.physics.uoc.gr/publications}}.

\begin{figure}
    \centering
    \includegraphics[scale=0.6]{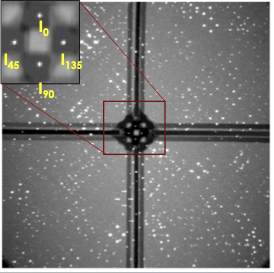}
    \caption{Image taken with RoboPol four-channel one-shot polarimetry system (ref: Ramaprakash et al., 2019\cite{robopol}). For each object, four images corresponding to the $0^{\circ}$, $45^{\circ}$, $90^{\circ}$ and $135^{\circ}$ polarizations are created by the WPs in close proximity on the CCD detector. Their differential photometry yields the Stokes parameters $q$ and $u$.}
    \label{robopol_image}
\end{figure}

RoboPol's polarization analyzer system consists of two quartz Wollaston Prisms (WPs), each with its own half-wave plate (HWP) at the front. The two WPs share a collimated beam created by upstream collimator optics. Four beams polarized along the $0^{\circ}$, $45^{\circ}$, $90^{\circ}$ and $135^{\circ}$ axes are split from this input beam and imaged as four spots in close proximity on the CCD detector, as shown in Figure~\ref{robopol_image}. Linear Stokes parameters $q$ and $u$ are obtained by performing relative photometry on the four images.

RoboPol is mounted on the direct Cassegrain focus of the telescope. While it has a large field of view of $13.6\times13.6~arcmin^{2}$, its performance is optimized inside the central region of $21\times21~arcsec^{2}$, which is masked from neighboring regions to reduce the sky background. For a detailed description of the instrument design, we refer readers to the RoboPol instrument design and commissioning paper by Ramaprakash et al.\cite{robopol}. The instrument was planned and designed for long-term high-accuracy and efficient monitoring of point sources such as blazars. Using measurements of on-sky polarized and unpolarized standard stars, the accuracy of RoboPol in degree of linear polarization, $p$, has been found to be between 0.1\% and 0.2\% in the R-band and marginally worse in other bands.

Besides its primary role in executing the core RoboPol scientific program, the RoboPol instrument also serves as a test-bed for the forthcoming \textsc{PASIPHAE} sky survey. This survey will utilize RoboPol's successor, the Wide-Area Linear Optical Polarimeter (WALOP) instruments. In preparation for the \textsc{PASIPHAE} program, RoboPol was instrumental in conducting two key projects: (a) a five-year monitoring initiative, which led to the creation of an extensive catalog of optical polarimetric standard stars\cite{RoboPol_standards}, and (b) the identification of new wide-field sources for calibrating wide-field instruments\cite{sky_polarization, dark_patches}.

In terms of instrument calibration and characterization, RoboPol is employed to test calibration methods for WALOP polarimeters. This paper focuses on one specific idea: enhancing the calibration and accuracy of four-channel one-shot polarimeters, such as RoboPol and WALOP, by utilizing a modulating HWP in conjugation with their one-shot polarimetry capacity.

To first order, and with accuracy sufficient for RoboPol, the relation of measured Stokes parameters $q_{m}$ and $u_{m}$   to the true Stokes parameters of a source $q$ and $u$  can be written in the form of Equations~\ref{calibration_eqn_q} and \ref{calibration_eqn_u}. Please refer to Appendix~\ref{instrument_matrix_appendix} for derivation of these equations and their relation to the instrument Mueller matrix. In general, these values can depend on all the intrinsic Stokes parameters of the source. The $b_{1}$ and $b_{2}$ terms are the \textit{polarimetric efficiencies} of the instrument, which capture how each of the measured Stokes parameters scales with respect to their corresponding input values. The $a_{1}$ and $a_{2}$ terms are referred to as the polarimetric \textit{zero offsets}, as they represent the measured Stokes parameters when the input is unpolarized. The terms $c_{1}$ and $c_{2}$ capture the instrument \textit{cross-talk} between linear Stokes parameters, i.e., how much of $u$ is converted into $q_m$, and $q$ into $u_m$, respectively. Likewise, $d_{1}$ and $d_{2}$ quantify the \textit{cross-talk} between the circular Stokes parameter and the linear Stokes parameters, i.e., how much of $v$ is converted into $q_{m}$ and $u_{m}$, respectively.

\begin{equation}\label{calibration_eqn_q}
    q_{m} = a_{1} + b_{1}q + c_{1}u  + d_{1}v
\end{equation}

\begin{equation}\label{calibration_eqn_u}
    u_{m} = a_{2} + b_{2}u + c_{2}q + d_{2}v  
\end{equation}

To estimate the coefficients, the instrument can be made to observe sources with known states of polarization, i.e., standard unpolarized and polarized stars for narrow field of view polarimeters. Based on RoboPol calibration measurements since 2013, it is found that $b_1$ and $b_2$ are 1 and $c_1$, $c_2$, $d_1$ and $d_2$ are 0.  Over the years, RoboPol has consistently shown an polarimetric zero-offset leading to linear polarization of 0.3\% to 0.4\% (please see Figure~8, Ramaprakash et al., 2019\cite{robopol}).
\par The non-zero value of $a_1$ and $a_2$, measured using unpolarized standard stars, arises from the preferential transmission of one of the orthogonal polarizations over the other in the optics downstream of the WP. This can be denoted by the often used \textit{correction factor}, $k = \frac{t_{o}}{t_{e}}$, where $t_{o}$ and $t_{e}$ are the (normalized) transmissions of the ordinary ($o$) and extraordinary ($e$) beams coming from either of the WPs. $k$ is related to the polarimetric zero-offsets as given by Equation~\ref{k_def}. Please note that non-zero values of $a_1$ and $a_2$ do not affect the instrument's polarimetric performance as long as they can be measured and corrected to the required accuracy levels. Thus, the source of limiting instrumental polarimetric accuracy is the scatter due to time-dependent variations in the $a_1$ and $a_2$ terms.

\begin{equation}\label{k_def}
a_i = \frac{t_o - t_e}{t_o + t_e} =   \frac{\frac{t_o}{t_e} - 1}{\frac{t_o}{t_e} + 1} = \frac{k_i - 1}{k_i + 1}
\end{equation}

\par The main limiting source of RoboPol’s current instrumental accuracy of 0.1~\% to 0.15~\% (and stability) is the time-varying instrument's transmission of the $o$ and $e$ beams downstream of the WP quantified by $k_i$ and thus $a_i$. It is anticipated that the cause for this is flexures at different telescope orientations and changes in ambient temperature at the observatory. These factors lead to changes in the relative optical path of the four beams and thus variable relative transmission. Therefore, in order to correct for these, a real-time system that can quantify this change in instrumental polarization is needed. Usually, any change in optical properties up-stream of the WP affects both the ordinary and extraordinary beams equally and gets cancelled.

Adding a modulating HWP either at the pupil before the WP or at the very beginning of the instrument near the focal plane allows for such a calibration system. For example, in conventional dual-channel polarimeters, a modulating HWP before the WP allows for measurements of $q$ and $u$ at HWP positions of $0^{\circ}$ and $22.5^{\circ}$. Additional measurements at $45^{\circ}$ and $67.5^{\circ}$ correspond to $-q$ and $-u$ measurements. Comparison of relative intensities of the $o$ and $e$ beam transmissions between $0^{\circ}$ and $45^{\circ}$ measurements and $22.5^{\circ}$ and $67.5^{\circ}$ measurements allow for estimation and correction for $a_1$ and $a_2$ in real-time using Equation~\ref{k_factor_def}, making them zero. In an ideal polarimeter, $0^{\circ}$ and $45^{\circ}$ measurements will flip the $I_o$ and $I_e$ intensities between them, and the deviation allows for polarimetric flat fielding and estimation of $k$ in real time. A similar method was used for improving the polarization accuracy of the WIRCPOL instrument (which is itself based on RoboPol design) from 1\% to 0.05\%\cite{WIRCPOL_calibrator}.

\begin{equation}\label{k_factor_def}
    k = \frac{I_{o}}{I_{e}} = [\frac{{I_{o}}^{0}\times{I_{o}}^{45}}{{I_{e}}^{0}\times{I_{e}}^{45}} ]^\frac{1}{2} \\
     = [\frac{{I_{o}}^{22.5}\times{I_{o}}^{67.5}}{{I_{e}}^{22.5}\times{I_{e}}^{67.5}} ]^\frac{1}{2}
     = [\frac{{I_{o}}^{0}\times{I_{o}}^{45}\times{I_{o}}^{22.5}\times{I_{o}}^{67.5}}{{I_{e}}^{0}\times{I_{e}}^{45}\times{I_{e}}^{22.5}\times{I_{e}}^{67.5}} ]^\frac{1}{4}
\end{equation}

Considering the RoboPol instrument as two independent dual-channel polarimeters, $k_1$ and $k_2$ denote the correction factors for the Left and Right halves of the RoboPol system. Nominally, the Left half measures the $q$ value through the normalized difference of $N_{0}$ and $N_{1}$ beams and the Right half measures the $u$ value through the normalized difference of $N_{2}$ and $N_{3}$ beams. The modulating HWP at the beginning of the instrument placed at angles that are multiples $45^{\circ}$ allows for estimating of $k_1$ and $k_2$ by ``polarimetric flat fielding"\cite{WIRCPOL_calibrator, impol}. In practise, we include additional exposures at HWP positions of $22.5^{\circ}$ and $67.5^{\circ}$ for enhanced robustness in estimation of $k_1$ and $k_2$ and for checking consistency of measurements, as given by Equations~\ref{k_1_def} and \ref{k_2_def}.
\begin{equation}\label{k_1_def}
    k_1 = [\frac{N_{0}^{0}\times{N_{0}}^{45}\times{N_{0}}^{22.5}\times{N_{0}}^{67.5}}{N_{1}^{0}\times{N_{1}}^{45}\times{N_{1}}^{22.5}\times{N_{1}}^{67.5}} ]^\frac{1}{4}
\end{equation}

\begin{equation}\label{k_2_def}
    k_2 = [\frac{N_{2}^{0}\times{N_{2}}^{45}\times{N_{2}}^{22.5}\times{N_{2}}^{67.5}}{N_{3}^{0}\times{N_{3}}^{45}\times{N_{3}}^{22.5}\times{N_{3}}^{67.5}} ]^\frac{1}{4}
\end{equation}
In this prescription, the total observation time of each target will be split into four exposures, each at calibration HWP positions of $0^{\circ}$, $22.5^{\circ}$, $45^{\circ}$, and $67.5^{\circ}$ degrees with respect to the instrument coordinate system (ICS). Thus, we get two values of $q$ and $u$ from each half of RoboPol corrected for the $k_i$ affect. Then these measurements of $q$  and $u$ are combined in arithmetic mean to provide the overall $q$ and $u$ values and associated variance in the measurements.

\par In this paper, we present the design of the RoboPol calibrator system and the on-sky performance obtained based on measurements of multiple standard stars in B, R, and I filters. Overall, we find that using the calibrator system, we obtain an accuracy better than 0.05~\% in the $q$ and $u$ measurements for all the three broadband filters, marking a factor of 2 to 3 improvement in instrumental sensitivity. 


\section{RoboPol Calibrator Design}

\par
\begin{figure}
    \centering
    \begin{subfigure}[b]{0.49\textwidth}
    \includegraphics[scale=0.074]{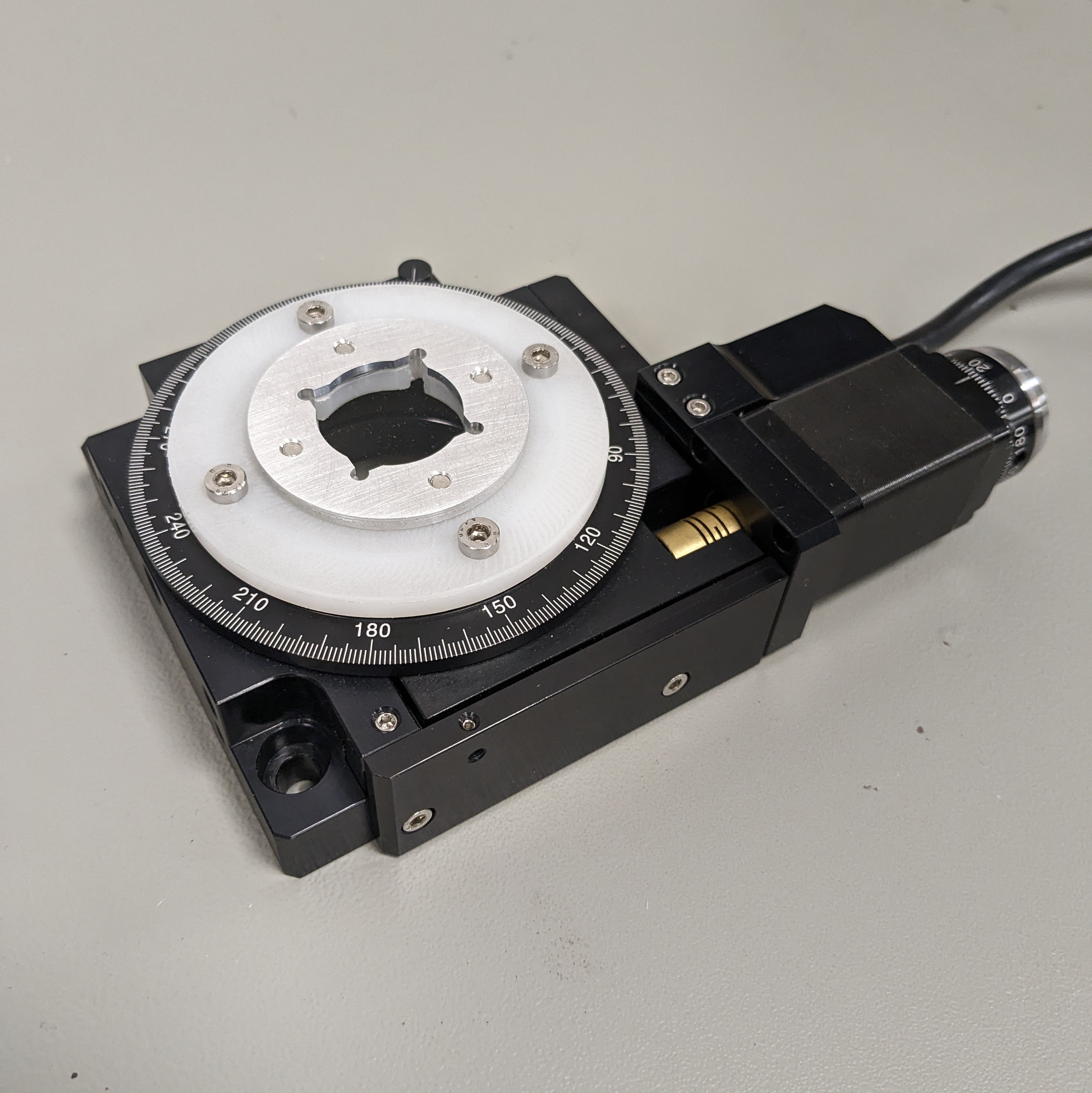}
    \caption{Real model of the RoboPol calibrator system. The HWP is mounted on a rotary stage.}
    \label{calibrator_stage}
    \end{subfigure}
    \begin{subfigure}[b]{0.49\textwidth}
    \includegraphics[scale=0.11]{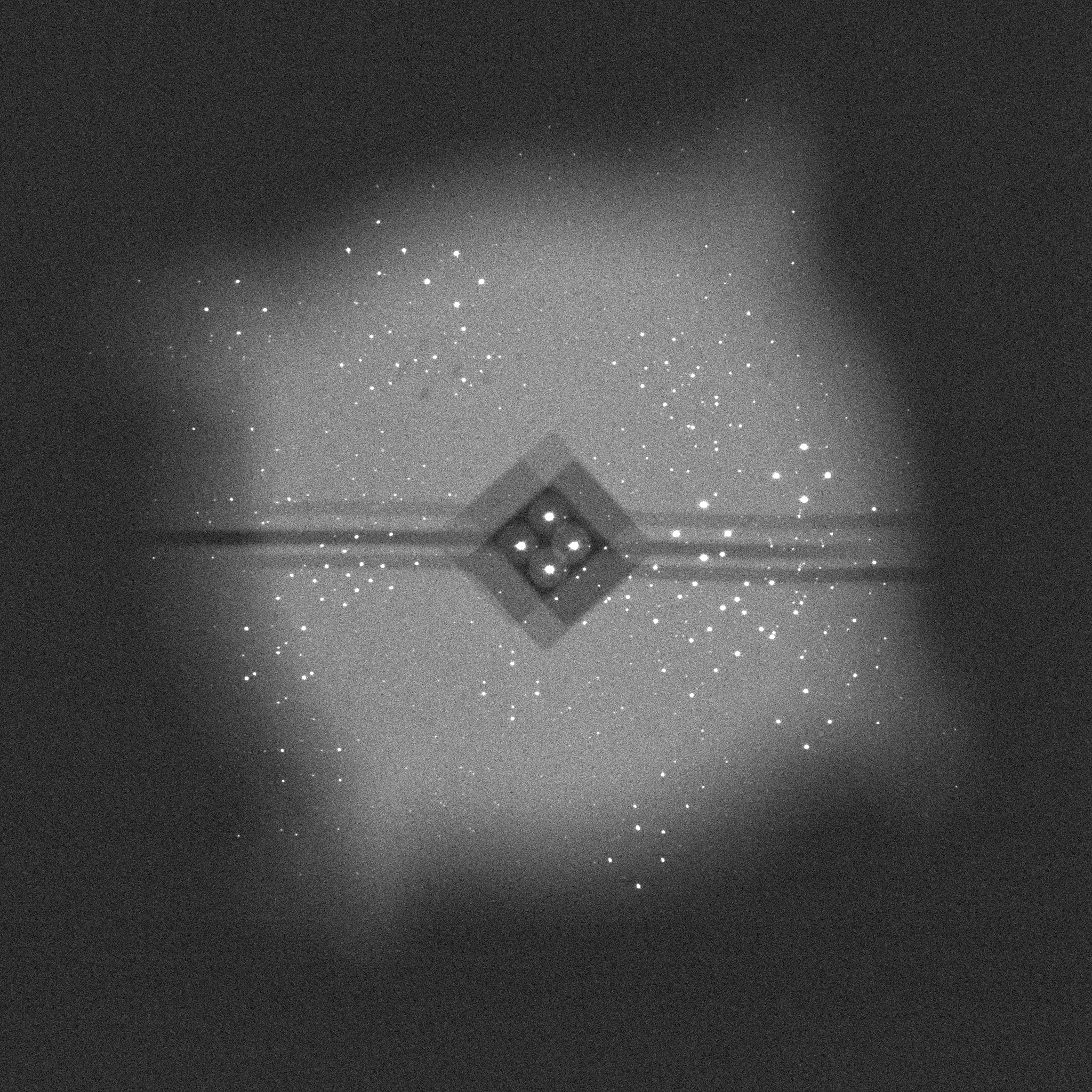}
    \caption{Image obtained through the calibrator system; the effective FoV is 7x7 arcminutes-squared.}
    \label{image_with_hwp}
    \end{subfigure}
    \caption{RoboPol Calibrator Design and image obtained using the calibrator system.}
    \label{}
\end{figure}

 The calibrator system for RoboPol presented in this paper consists of half-wave plate (HWP) made of quartz and MgF$_{2}$ crystals that is then mounted on a Standa rotary stage (Part\# 8MR190-2-28). The HWP is optimized for performance in the optical wavelengths of $400-800~nm$, which encompasses all the broadband filters of RoboPol. The mounting adapters to place the HWP on the rotary stage is either 3-D printed or made in local machining workshops in Crete. A picture of the calibrator system is shown in Figure~\ref{calibrator_stage}. The rotary stage is controlled by a Controller Box developed in-house by the lab at Inter-University Centre for Astronomy and Astrophysics (IUCAA), Pune, India. The calibrator system is mounted on top of the RoboPol instrument, just ahead of the focal plane/mask plane. The aperture of the HWP used is 25~mm$~\times$~25~mm, smaller than the nominal FoV size at the telescope focal plane. The size of the HWP in combination with the mounting system vignettes light from regions at outer part of the original RoboPol FoV. The obtained FoV with the HWP calibrator is roughly $7\times7~arcmin^{2}$; an image obtained from RoboPol with the calibrator system mounted on it is shown in Figure~\ref{image_with_hwp}.


\section{Observations and Analysis}
\par To test the on-sky performance of the calibrator system, it was mounted on RoboPol, and observations were carried out for a sample of unpolarized and polarized standard stars traditionally used for RoboPol observations.  Table~\ref{unpol_measurements_table} and \ref{pol_measurements_table} list all the standard star measurements conducted for this project. To estimate the long-term performance with this system, measurements were taken on 5 nights spread over 2 RoboPol observing seasons.
As noted earlier, each observation consists of one or multiple sets of four exposures with RoboPol at HWP positions of $0^{\circ}$, $22.5^{\circ}$, $45^{\circ}$, and $67.5^{\circ}$ with respect to the ICS. 

For the analysis of the raw data, an automated data reduction pipeline was developed using the Python programming language. Attention was given towards error estimation and propagation in each step of the analysis. Photometry was performed using the aperture photometry package of the Photutils software\footnote{\url{https://photutils.readthedocs.io/en/stable/}}. Curve of growth plots were created for each image, and appropriate aperture and annulus radii were chosen. Using this procedure, the intensities $I_{o}$ and $I_e$ were obtained for all HWP positions. Differential transmission for the $o$ and $e$ beams, characterized by the correction factors $k_1$ and $k_2$ (as given by Equations~\ref{k_1_def} and \ref{k_2_def}), was calculated. Subsequently, the 
$o$ beam intensity was corrected as ${I_{o}^{'}}= \frac{I_{o}}{k_i} $. The normalized Stokes number is then given by the normalized difference between the two intensities, as given by Equation~\ref{rm}.

$R_l(\alpha)$ and $R_r(\alpha)$ are the Stokes parameter values from the two Halfs of RoboPol that corresponds to either $\pm~q$ or $\pm~u$, depending on the HWP position $\alpha$, as given by Equations~\ref{R_eqn} and \ref{R_eqn2}, where the subscripts $l$ and $r$ correspond to the left and right Half measurements. The overall $q$ and $u$ are found using the Equations~\ref{q_NOT} and \ref{u_NOT} and the associated uncertainty in the measurements is found by taking the standard deviation of the corresponding $r_m$ values.


\begin{equation}\label{rm}
{R} = \frac{{I_{o}^{'}}- {I_{e}}}{{I_{o}^{'}}+ {I_{e}}}
\end{equation}

\begin{equation}\label{R_eqn}
    R_l(\alpha) = \rm{q}\times cos4\alpha + u\times sin4\alpha
\end{equation}
\begin{equation}\label{R_eqn2}
    R_r(\alpha) = \rm{-q}\times sin4\alpha + u\times cos4\alpha
\end{equation}
\begin{equation}\label{q_NOT}
q = \frac{R_l(0) - {R_l}(45)- R_r(22.5) + R_r(67.5)}{4}
\end{equation}

\begin{equation}\label{u_NOT}
u = \frac{R_l(22.5) - {R_l}(67.5) + R_r(0) - R_r(45)}{4}
\end{equation}

\begin{table}
\centering
\begin{tabular}{lcr}
\hline\\
\textbf{Source Id}               & \textbf{Filter Used}   & \textbf{Date of Observation} \\
\hline
\multirow{4}{*}{BD+32.3739} &      \multirow{4}{*}{R}             & Jun 8, 2022                   \\

                              &                   & Jul 1, 2023                   \\
                              &                   & Jul 2, 2023                   \\
                              &                   & Nov 15, 2023                  \\ \hline
\multirow{2}{*}{BD+32.3739} & \multirow{2}{*}{R} & Jul 1, 2023                   \\
                              &                   & Jul 2, 2023                   \\ \hline
\multirow{4}{*}{BD+32.3739} & \multirow{4}{*}{B} & Jun 8, 2022                   \\
                              &                   & Jul 1, 2023                   \\
                              &                   & Jul 2, 2023                   \\
                              &                   & Nov 15, 2023                  \\ \hline
\multirow{2}{*}{BD+32.3739} & \multirow{2}{*}{B} & Jul 1, 2023                   \\
                              &                   & Jul 2, 2023                   \\ \hline
\multirow{2}{*}{BD+32.3739} & \multirow{2}{*}{I} & Jul 1, 2023                   \\
                              &                   & Jul 2, 2023                   \\ \hline
\multirow{3}{*}{BD+28.4211 } & \multirow{3}{*}{R} & Jul 1, 2023                   \\
                              &                   & Nov 15, 2023                 \\
                              &                   & Nov 15, 2023                 \\ \hline
\multirow{2}{*}{BD+28.4211} & \multirow{2}{*}{B} & Nov 15, 2023                 \\
                              &                   & Nov 15, 2023                 \\ \hline
BD+40.2704                   &          R         & Jun 8, 2022                   \\ \hline
BD+40.2704                   &                   & Jun 8, 2022                   \\ \hline
\multirow{2}{*}{HD212311}   & \multirow{2}{*}{R} & Jul 1, 2023                   \\
                              &                   & Nov 15, 2023                  \\ \hline
HD212311                    &           B        & Nov 15, 2023                  \\ \hline
\end{tabular}
\caption{List of all observations for unpolarized standards.}
\label{unpol_measurements_table}
\end{table}

\begin{table}[]
\centering
\begin{tabular}{ccccc}
\hline
\textbf{Source Id}                     & \textbf{Filter Used} & \textbf{Date of Observation}  & \begin{tabular}[c]{@{}c@{}}\textbf{Catalog Value}\\ p \%\end{tabular} & \begin{tabular}[c]{@{}c@{}}\textbf{Catalog Value}\\ EVPA (in Deg)\end{tabular} \\
\hline
HD155197           &     R  & Jul 1, 2023 & 4.986                                               & 102.88                                                                \\ \hline
HD155528                     &    R   & Jul 1, 2023          & {5.21}                                                & 92.61                                                                 \\ \hline
\multirow{3}{*}{Hiltner960 } &   \multirow{3}{*}{R}    & Jul 1, 2023          & {5.21}                                                & 54.54                                                                 \\
                              &        & Jul 2, 2023          & {5.21}                                                & 54.54                                                                 \\
                              &        & Nov 15, 2023         & {5.8}                                                 & 54.54                                                                 \\ \hline
\multirow{2}{*}{HD183143}   &    \multirow{2}{*}{R}   & Jun 8, 2022          & \multirow{2}{*}{{5.8}}                                & \multirow{2}{*}{178.6}                                                                 \\
                              &        & Nov 15, 2023         &                                                              &                                                                 \\ \hline
\multirow{2}{*}{HD183143}   &   \multirow{2}{*}{B}    & Jun 8, 2022          &   \multirow{2}{*}{{5.8}}                               & \multirow{2}{*}{178.6}                                                                 \\
                              &        & Nov 15, 2023         &                                                              &                                                                       \\ \hline
HD183143                     &   I    & Jun 8, 2022          & {3.13}                                                & 178.6                                                                 \\ \hline
\multirow{2}{*}{CygOB2}     &    \multirow{2}{*}{ R}   & Jun 8, 2022          & \multirow{2}{*}{3.13}                                                &\multirow{2}{*}{86}                                                                   \\
                              &       & Nov 15, 2023         &                                                              &                                                                     \\ \hline
CygOB2                       &    R   & Nov 15, 2023         & {4.893}                                               & 86                                                                    \\ \hline
HD204827                    &     R  & Nov 15, 2023         & {5.648}                                               & {59.1}                                                         \\ \hline
HD204827                     &    B   & Nov 15, 2023         &          {5.648}                                                     & 58.2                                                         \\ \hline
\end{tabular}
\caption{List of all observations for polarized standards.}
\label{pol_measurements_table}
\end{table}

\begin{figure}
    \centering
    \includegraphics[scale= 0.75]{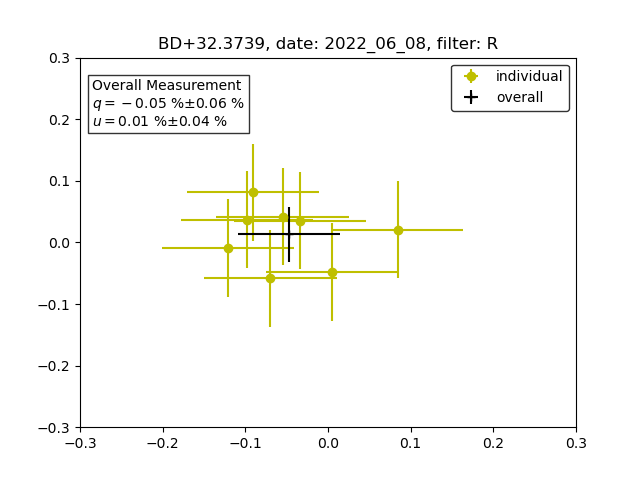}
    \caption{Plot of measured $q$ and $u$ values of the unpolarized standard star BD+32.3739
(yellow crosses) along with the average value and the observed standard deviation (black cross).
The top left legend lists the overall values. The median individual errors in yellow are that of the expected errors resulting from photon noise.}
    
    \label{single_object}
\end{figure}






\section{Results}\label{results_sec}

Figure~\ref{single_object} is a plot of the polarimetric measurement of a single unpolarized standard star BD+32.3739 in the SDSS-r filter. As noted above, we get 2 measurements of $q$ and $u$ from each half of RoboPol, thus 4 $q-u$ values in total per cycle. For the observation in Figure~\ref{single_object}, 2 sets of such measurements were done. From the mean and variance, we get the overall polarization of the source and the associated standard deviation. As can be noted, the final values of $q$ and $u$ are completely corrected for instrumental polarization through the modulating effect of the calibrator HWP; there is no residual measured polarization within the values of the spread of measurements ($<$~0.05~\%). 

\begin{figure}
    \centering
\begin{subfigure}{0.99\textwidth}
    \centering
    \includegraphics[scale=0.55]{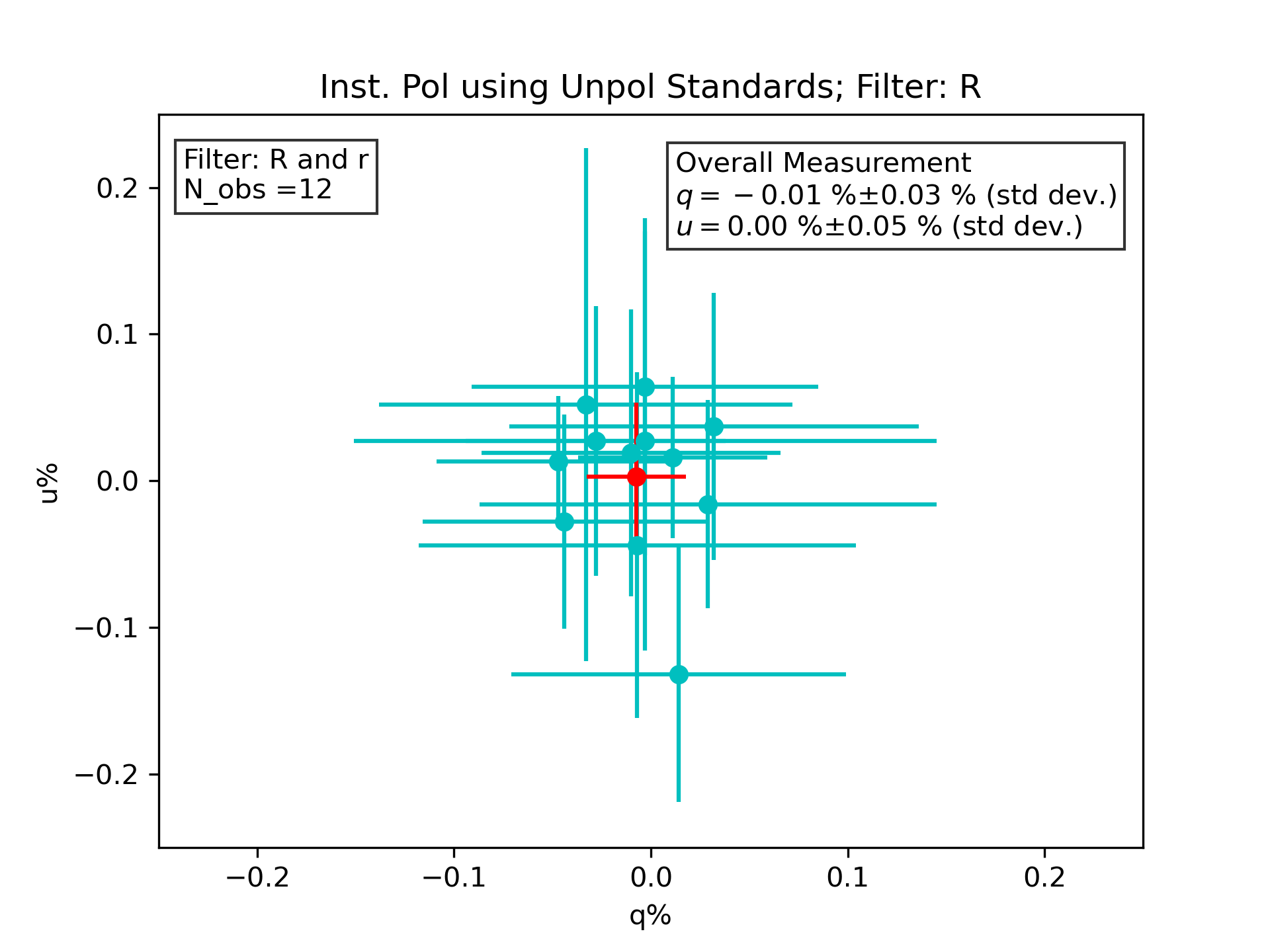}
    \caption{R band}
\end{subfigure}
\begin{subfigure}{0.49\textwidth}
    \centering
    \includegraphics[scale=0.55]{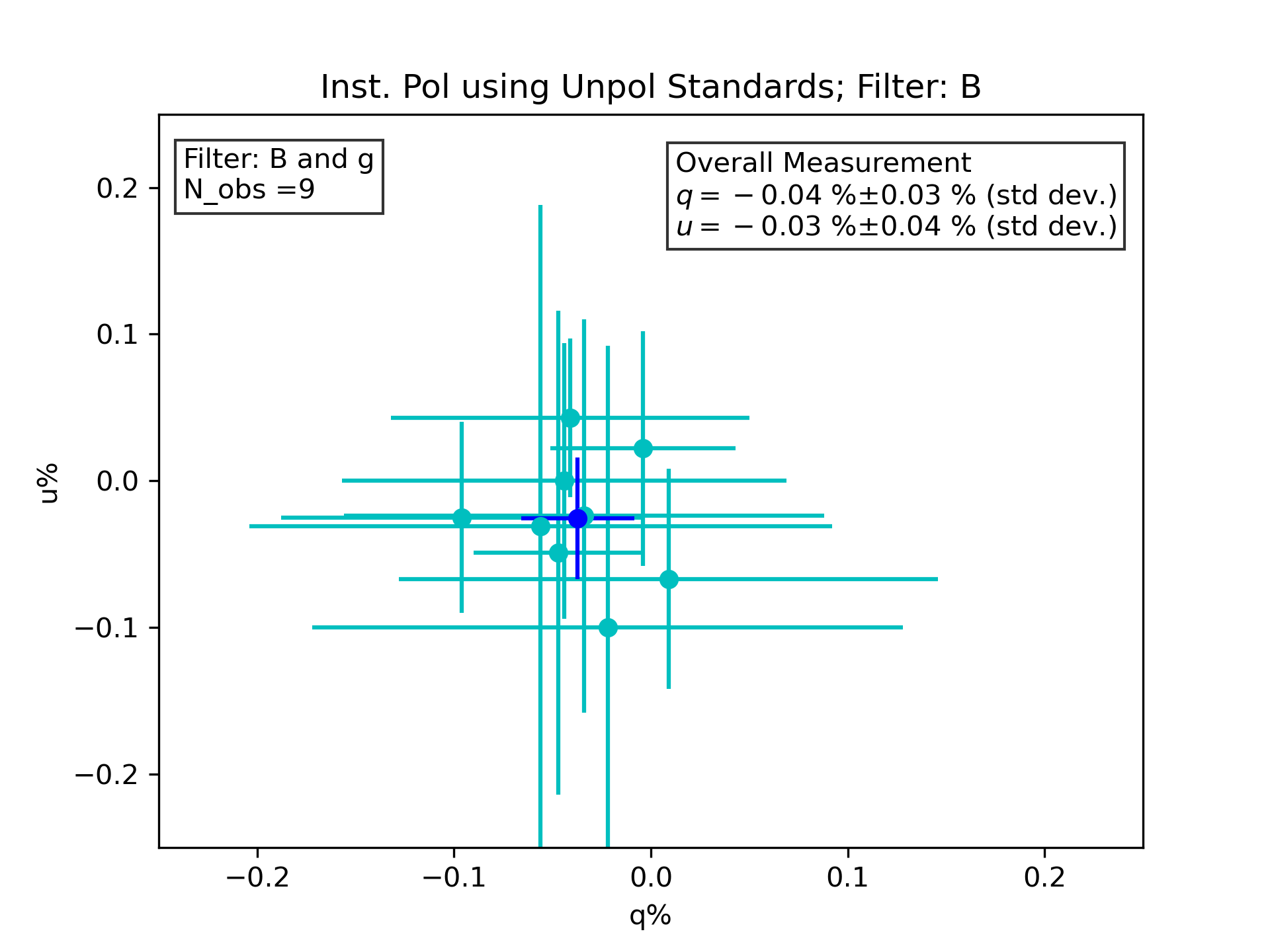}
    \caption{B band}
\end{subfigure}
\begin{subfigure}{0.49\textwidth}
    \centering
    \includegraphics[scale=0.55]{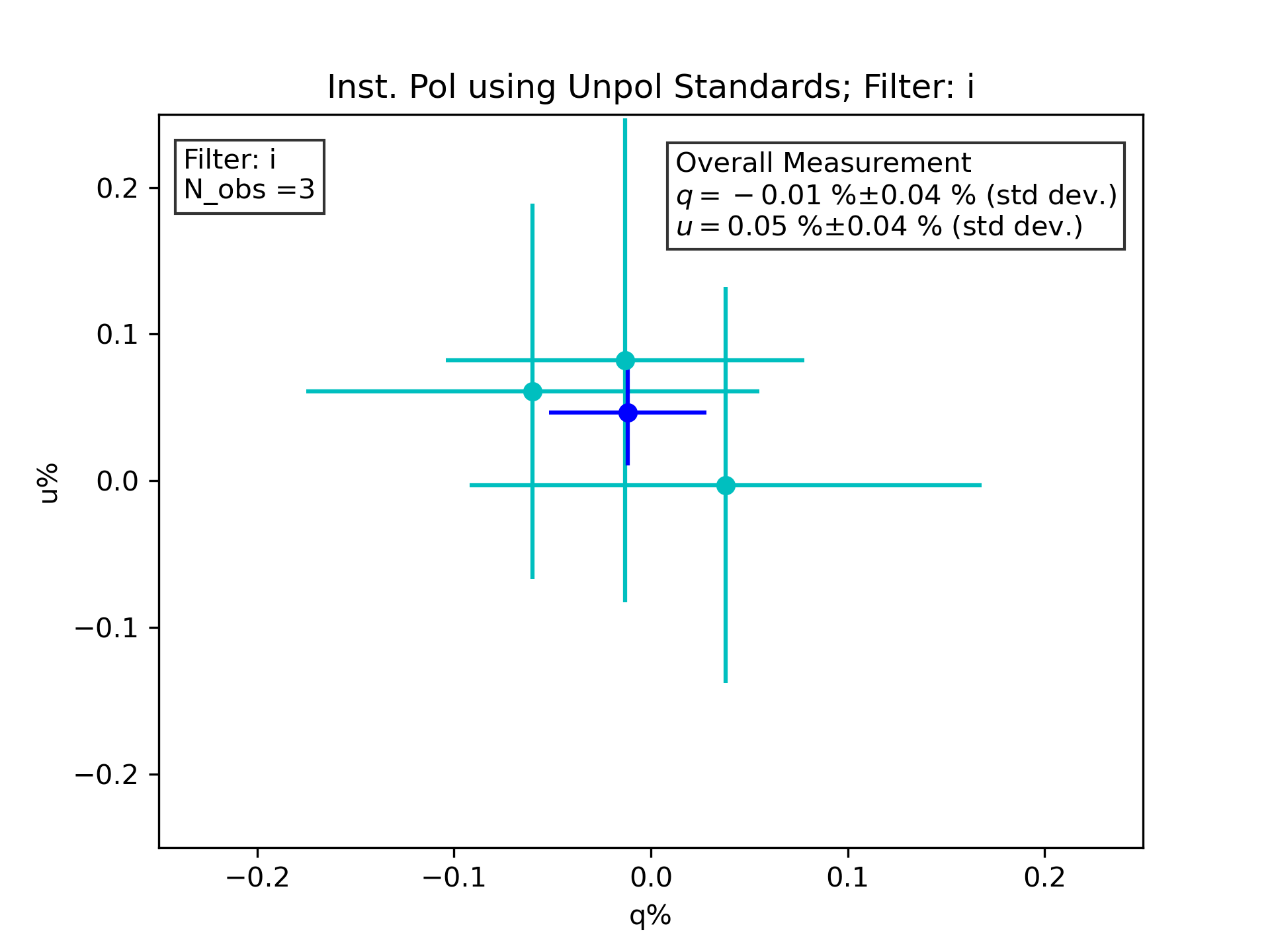}
    \caption{I band}
\end{subfigure}
    \caption{Instrumental polarization using unpolarized standards in R, B and I bands when aggregated over all observations across two RoboPol seasons.}
    \label{overall_zero_accuracy}
\end{figure}

To quantify the long-term accuracy and sensitivity of the instrument with the calibrator, measurements of all the unpolarized standards stars are collated for each filter. Without segregating as a function of date or season, the overall results are shown in Figure~\ref{overall_zero_accuracy}. As can be seen through the plots, for each filter, the accuracy in $q$ and $u$ is better than 0.05~\% in both $q$ and $u$. Equally importantly, the absolute value of instrumental polarization is zero within measurement uncertainties. Please note that in this accuracy regime, a small or significant part of the variance seen in the measurements will be originating from the variability and non-zero value of the standard unpolarized stars used for calibration themselves.

\begin{figure}
    \centering
\begin{subfigure}{0.99\textwidth}
    \centering
    \includegraphics[scale=0.6]{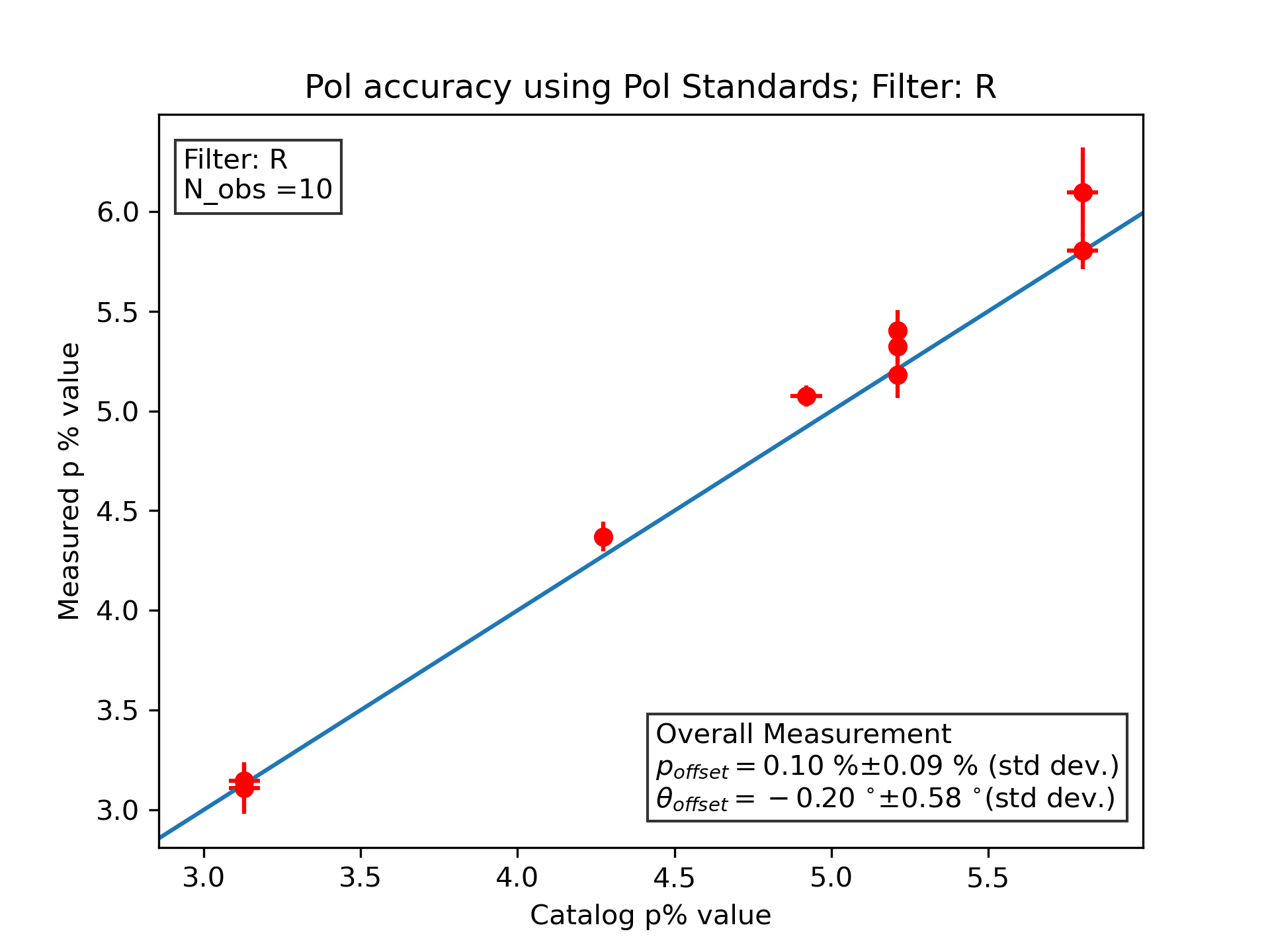}
    \caption{R band}
    \label{fig:enter-label}
\end{subfigure}
\begin{subfigure}{0.49\textwidth}
    \centering
    \includegraphics[scale=0.55]{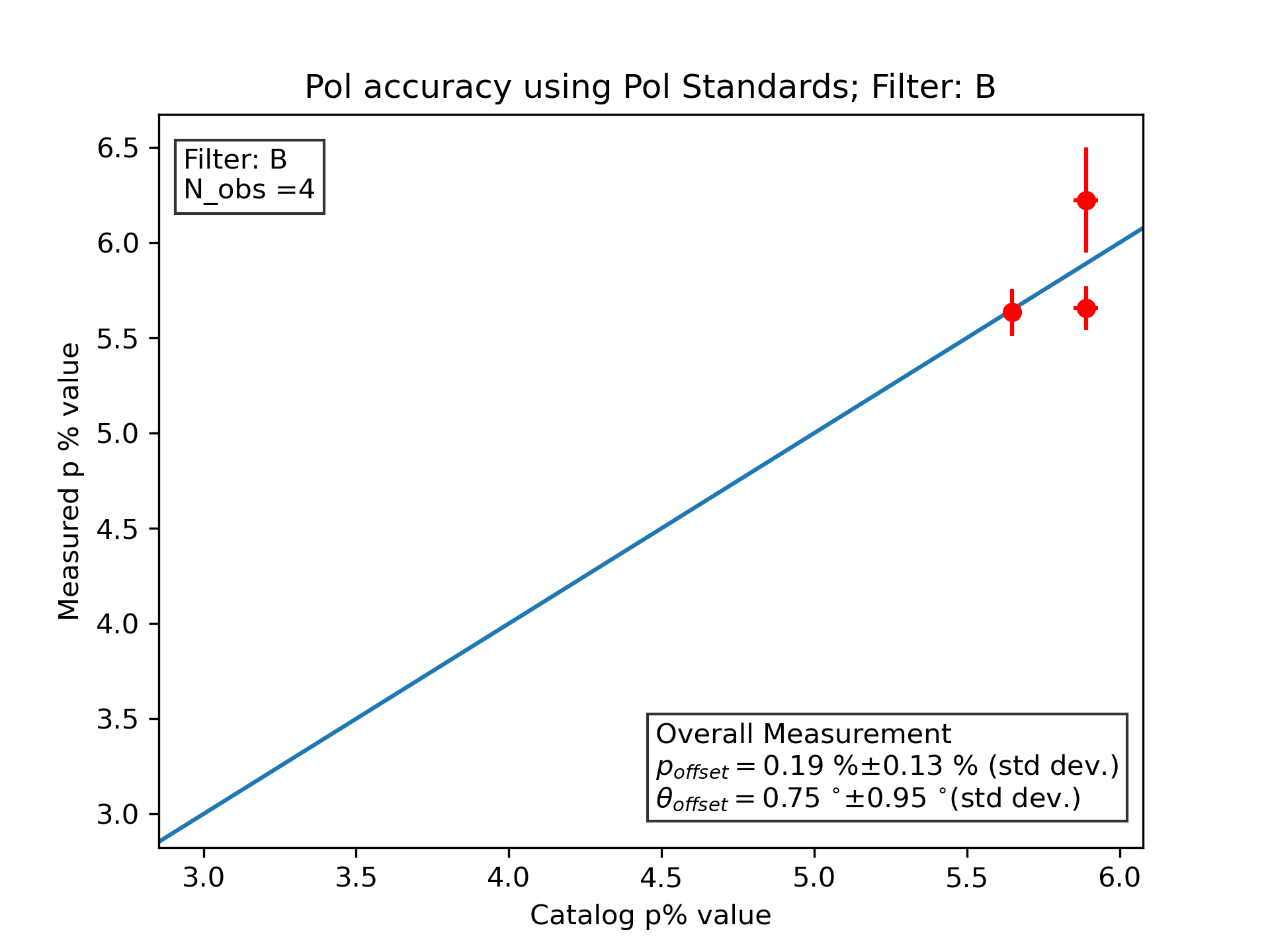}
    \caption{B band}
    \label{fig:enter-label}
\end{subfigure}
\begin{subfigure}{0.49\textwidth}
    \centering
    \includegraphics[scale=0.55]{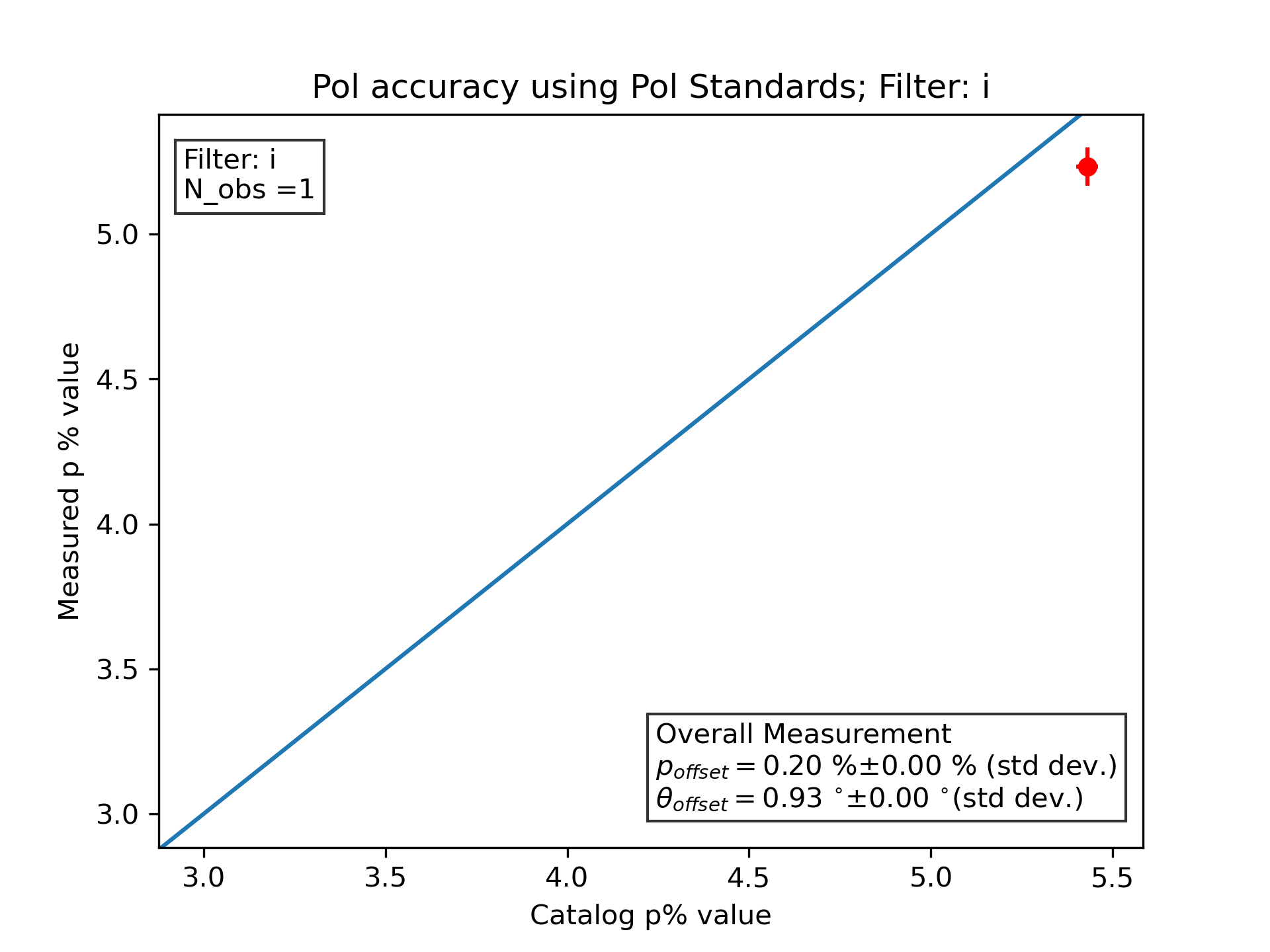}
    \caption{I band}
    \label{fig:enter-label}
\end{subfigure}
    \caption{Polarimetric accuracy and efficiency using polarized standards in R, B and I bands. In the legend, the overall comparison of measured values to catalog values is noted.}
    \label{overall_pol_accuracy}
\end{figure}

Figure~\ref{overall_pol_accuracy} plots the instrument-measured polarization values, $p$, in comparison to the standard polarized stars used in the polarimetry community. In the R band, it can be noted that the measured fractional polarization values match the catalog values to better than 0.1~\% accuracy. Furthermore, the measured polarization angle matches the catalog value to better than $0.5^{\circ}$ in the R band after correction for the instrument's rotation. This represents an improvement by a factor of 2. More measurements are needed in the B and I bands to make a robust claim for these filters.

\section{Conclusions}

\par Table~\ref{comparison} compares the performance of the new calibrator system with the normal RoboPol mode. There are two key improvements. Firstly, in just the R band, there is at least a two-fold improvement in the accuracy of the instrument ($\sigma_p$), as well as better matching with catalog polarization values of standard stars. Secondly, the HWP calibrator improves the accuracy in all broadband filters, which will enable RoboPol to carry out new scientific studies that require high accuracy multi-filter polarimetric studies. An example of such research includes the measurement of Serkowski curves along the ultra-diffuse interstellar medium sightlines, where dust-induced polarization is typically very low, at levels of $0.1\%$ or lower. Prior to the implementation of the HWP calibrator, RoboPol was limited in its ability to accurately measure this phenomenon, often only able to provide upper limits on the signal due to constraints imposed by instrument accuracy.

\begin{table}[]
    \centering
    \begin{tabular}{ccc}
    \hline
    Filter& Instrumental polarization with calibrator & Instrumental polarization without calibrator\\
      & $p_{inst} \pm {\sigma}_{p}$ & $p_{inst} \pm {\sigma}_{p}$\\
    \hline
     B  & 0.05\% $\pm$ 0.03\%  &  0.29\%~ $\pm$~ 0.16\%  \\
     R  & 0.01\% $\pm$ 0.03\%  &  0.30\%~ $\pm$~ 0.09\%   \\
     I  & 0.05\% $\pm$ 0.04\%  &  0.60\%~ $\pm$~ 0.08\%\\
       \hline
    \end{tabular}
    \caption{Comparison of RoboPol instrumental polarization and accuracy after polarimetric calibrations estimated through observations of multiple unpolarized standard stars with the HWP calibrator and in conventional mode without the calibrator. The values for the instrument accuracy in conventional mode are taken from the main RoboPol instrument paper by Ramaprakash et al., 2019\cite{robopol}.}
    \label{comparison}
\end{table}

\acknowledgments 

The PASIPHAE program is supported by grants from the European Research Council (ERC) under grant agreement No 771282 and No 772253, from the National Science Foundation, under grant number AST-1611547 and AST-2109127, and the National Research Foundation of South Africa under the National Equipment Programme. This project is also funded by an infrastructure development grant from the Stavros Niarchos Foundation and from the Infosys Foundation. K.T. acknowledges support from the Foundation for Research and Technology – Hellas Synergy Grants Program through project POLAR, jointly implemented by the Institute of Astrophysics and the Institute of Computer Science. 
\par We thank Anna Steiakaki for her help in various phases of the project, including component fabrication and installation on the telescope. 

\appendix
\section{Measured Stokes Parameters and Instrument Matrix}\label{instrument_matrix_appendix}
The incident intensity at the detector for any polarization channel of the instrument ($0^\circ$, $45^\circ$, $90^\circ$ and $135^\circ$ polarizations) can be written as:
    \begin{equation}
    I_{\theta} = a_{\theta} + b_{\theta}q + c_{\theta}u 
\end{equation}
where $a_{\theta}$, $b_{\theta}$ and $c_{\theta}$ correspond to the elements of the first row of the Mueller matrix for the optical path corresponding to that polarization angle/channel. The normalized difference between intensities corresponding to two orthogonal polarization angles/channels yields $q$ or $u$, as given by the following equation.
\begin{equation}
    r_{i} = \frac{I_{\theta1} - I_{\theta2}}{I_{\theta1} + I_{\theta2}}
    = \frac{(a_{\theta1} + b_{\theta1}q + c_{\theta1}u) - (a_{\theta2} + b_{\theta2}q + c_{\theta2}u)} {(a_{\theta1} + b_{\theta1}q + c_{\theta1}u) + (a_{\theta2} + b_{\theta2}q + c_{\theta2}u)} 
\end{equation}
The binomial expansion of $(1+x)^{-1}$ is given by $(1+x)^{-1} = 1 - x +x^2 -x^3 + x^4$. Using that, the generalized normalized difference can be written as a polynomial equation in $q$ and $u$.
\begin{equation}
 r_{i} = A_{i} + B_{i}q + C_{i}u + D_{i}q^2 + E_{i}u^2 + F_{i}qu + ...
\end{equation}
In general, for most simple polarimeters, the second order terms are zero, and the instrument measured Stokes parameters can be written as a set of linear equations, together forming a Muller matrix like \textit{Instrument Matrix}. The first row is inconsequential to polarimetric measurements and hence can be ignored.
\begin{gather}
s_{m} = 
\begin{bmatrix}
1 \\
q_{m} \\
u_{m} \\
v_{m} 
\end{bmatrix}
= 
m_{inst}\times s
= 
\begin{bmatrix}
m_{11} & m_{12} & m_{13} & m_{14}  \\
m_{21} & m_{22} & m_{23} & m_{24}  \\
m_{31} & m_{32} & m_{33} & m_{34}  \\
m_{41} & m_{42} & m_{43} & m_{44} 
\end{bmatrix}
\times  
\begin{bmatrix}
1 \\
q \\
u \\
v 
\end{bmatrix}
\\
=
\begin{bmatrix}
- & - & - & -  \\
1 \,\to\,q_{m} & q\,\to\,q_{m} & u\,\to\,q_{m} & v\,\to\,q_{m}  \\
1 \,\to\,u_{m} & q\,\to\,u_{m} & u\,\to\,u_{m} & v\,\to\,u_{m}  \\
1 \,\to\,v_{m} & q\,\to\,v_{m} & u\,\to\,v_{m} & v\,\to\,v_{m}
\end{bmatrix}
\times  
\begin{bmatrix}
1 \\
q \\
u \\
v 
\end{bmatrix}
\end{gather}

\bibliography{article} 
\bibliographystyle{spiebib} 

\end{document}